\documentclass[10pt]{article}
\scrollmode
\usepackage{amsmath,amssymb}
\usepackage{eucal}
\usepackage[dviwindo]{graphicx}
\usepackage[dviwindo]{graphics}

\numberwithin{equation}{section}
\newcommand{\ben}{\begin{eqnarray}}
\newcommand{\een}{\end{eqnarray}}
\newcommand{\la}{\label}

\begin{document}

\title{Novel relations and new properties of confluent Heun's functions and their derivatives of arbitrary order}

\vskip 1.5truecm

\author{Plamen~P.~Fiziev\thanks{Department of Theoretical Physics, University of Sofia,
Boulevard 5 James Bourchier, Sofia 1164, Bulgaria,
E-mail:\,\,\,fiziev@phys.uni-sofia.bg}}

\date{}
\maketitle

\begin{abstract}

The present article reveals important properties of the confluent Heun's functions.
We derive a set of novel relations for confluent Heun's functions and their derivatives of arbitrary
order. Specific new subclasses of confluent Heun's functions are introduced and studied.
A new alternative derivation of confluent Heun's polynomials is presented.
\end{abstract}

\hskip 1truecm PACS: 02.30.Gp, 02.30.Hq



%
\sloppy
\scrollmode

\section{Introduction}

The solutions of the confluent Heun's
differential equation \cite{Heun}-\cite{SL}, written in the simplest uniform shape \cite{PF2009a}
\begin{align}
H''+\left(\alpha+{\frac{\beta+1}{z}}+{\frac{\gamma+1}{z-1}}\right)H'+
\left( {\frac\mu z}+{\frac\nu {z-1}} \right)H = 0,\la{DHeunC}
\end{align}
are of continuous and significant interest for many applications in different areas of natural sciences and especially in physics.

The equation \eqref{DHeunC} has three singular points: two regular ones -- $z=0$ and $z=1$, and one irregular one -- $z=\infty$.
The standard confluent Heun's function $\text{HeunC}(\alpha,\beta,\gamma,\delta,\eta,z)$
is a unique particular solution, which is regular around the regular singular point $z=0$.
It is defined via the convergent in the disk $|z|<1$ Taylor series expansion
\ben
\text{HeunC}(\alpha,\beta,\gamma,\delta,\eta,z)=\sum_{n=0}^\infty v_{n}(\alpha,\beta,\gamma,\delta,\eta)z^n,
\la{HeunC}
\een
assuming the normalization $\text{HeunC}(\alpha,\beta,\gamma,\delta,\eta,0)=1$.
The parameters $\alpha, \beta, \gamma, \delta, \eta $, introduced in \cite{DDMRR, DMR} and used in the widespread computer package Maple,
are related with $\mu$ and $\nu$ according to the equations
$\mu ={\frac{1} 2}(\alpha-\beta -\gamma+\alpha\beta-\beta\gamma)-\eta$ and
$\nu = {\frac{1} 2}(\alpha+\beta +\gamma+\alpha\gamma+\beta\gamma)+\delta + \eta$.

The coefficients $v_{n}(\alpha,\beta,\gamma,\delta,\eta)$ are determined by three-term recurrence relation
\ben
A_{n}v_{n}=B_{n}v_{n-1}+C_{n}v_{n-2}
\la{recurrence}
\een
with initial condition $v_{-1}=0,\,\,v_{0}=1$. Here
\ben
\hskip -.truecm A_{n}&=&1+{\frac{\beta}{n}}\,\to 1,\,\,\,\text{when}\,\,\,n\to\infty, \nonumber\\
\hskip -.truecm B_{n}&=&1+{\frac{-\alpha+\beta+\gamma-1}{n}}+{\frac{\eta-(-\alpha+\beta+\gamma)/2-\alpha\beta/2+\beta\gamma/2}{n^2}}\,\to 1,\,\,\,\text{when}\,\,\,n\to\infty,\nonumber\\
\hskip -.truecm C_{n}&=&{\frac{\alpha}{n^2}}\left({\frac \delta \alpha}+{\frac {\beta+\gamma}{2}}+n-1\right)\,\to 0,\,\,\,\text{when}\,\,\,n\to\infty.
\la{rec_coeff}
\een

According to \cite{Heun}-\cite{SL} the function $\text{HeunC}(\alpha,\beta,\gamma,\delta,\eta,z)$
reduces to a polynomial of degree $N\geq 0$ with respect to the variable $z$
if and only if the following two conditions are satisfied:
\begin{subequations}
\la{PolynomCond:ab}
\ben
{\frac{\delta}{\alpha}}+{\frac{\beta+\gamma}{2}}+N+1=0,\la{PolynomCond:a}\\
\Delta_{N+1}(\mu)=0.\la{PolynomCond:b}
\een
\end{subequations}
Further on we call the first condition \eqref{PolynomCond:a}
-- a "$\delta_N$-condition", and the second one \eqref{PolynomCond:b} -- a "$\Delta_{N+1}$-condition".
One can find an explicit form of the left hand side $\Delta_{N+1}(\mu)$
of the condition \eqref{PolynomCond:b}, convenient for practical calculations, in the Appendix.

Indeed, the $\delta_N$-condition is equivalent to the equation $C_{N+2}=0$,
and the $\Delta_{N+1}$-condition turns to be equivalent to
the requirement $v_{N+1}(\alpha,\beta,\gamma,\delta,\eta)=0$.
Then as a result of equation \eqref{recurrence} and additional conditions \eqref{PolynomCond:ab}
we obtain $v_{N+2}(\alpha,\beta,\gamma,\delta,\eta)=0$.
Since two consecutive terms in the tree-term recurrence relation (\ref{recurrence}) are zero,
all next terms are zero, too.
Hence, under simultaneous fulfillment of the two additional conditions \eqref{PolynomCond:ab}
the confluent Heun function $\text{HeunC}(\alpha,\beta,\gamma,\delta,\eta,z)$
(\ref{HeunC}) reduces to a polynomial of degree $N$.

In present article we derive a set of novel relations and differential equations
for confluent Heun's functions \eqref{HeunC} and their derivatives
${\frac{d^n}{dz^n}}\text{HeunC}(\alpha,\beta,\gamma,\delta,\eta,z)$ of arbitrary order $n$
-- see Section 2.
In Section 3 we introduce a new subclass of confluent  Heun's functions
$\text{HeunC}_N(\alpha,\beta,\gamma,\eta,z)$, which obey only the $\delta_N$-condition
\eqref{PolynomCond:a}, as well as their associate confluent Heun's functions
$\text{HeunC}^\maltese_N(\alpha,\beta,\gamma,\eta,z)$.
In Section 4 we utilize the newly found relations to present an alternative derivation of
the confluent Heun's polynomials without use of the recurrence relation \eqref{recurrence}.
Thus our consideration reveals important new properties of the confluent Heun functions.

In the concluding Section 5 we briefly discuss the relation of our results for confluent
Heun's functions with similar results for general Heun's functions.

The confluent Heun's functions $\text{HeunC}(\alpha,\beta,\gamma,\delta,\eta,z)$,
the specific classes of $\delta_N$-confluent Heun's functions
$\text{HeunC}_N(\alpha,\beta,\gamma,\eta,z)$ and their associated functions
$\text{HeunC}^\maltese_N(\alpha,\beta,\gamma,\eta,z)$, as well as confluent Heun's polynomials
$\text{PHeunC}_{N,k}(\alpha,\beta,\gamma,z)$ play a very important role in some applications,
especially in gravitational physics \cite{PF2009a}-\cite{BS}. The present study of their properties
was inspired by our desire to reach a true mathematical and physical understanding
of the important Teukolsky-Starobinsky identities,
derived and used in the early articles \cite{PressTeukolsky1973}-\cite{Chandra1984}.
We present here the formal mathematical results,
which can be applied, too, in other scientific domains,
both for analytical and for numerical calculations.

\section{Novel relations for the confluent Heun's functions and their derivatives}

Let us define the differential expression
\begin{align}
\hat D_{\alpha,\beta,\gamma,\delta,\eta}=z(z-1)
\left({\frac{d^2}{dz^2}}+\Big(\alpha+{\frac{\beta+1}{z}}+{\frac{\gamma+1}{z-1}}\Big){\frac{d}{dz}}+
{\frac \mu  z}+{\frac\nu {z-1}} \right).\la{D}
\end{align}
One can use it to write down the confluent Heun equation (\ref{DHeunC}) in the following compact form
\ben
\hat D_{\alpha,\beta,\gamma,\delta,\eta} H=0.
\la{DHeunC_}
\een
It can be shown that certain restrictions of differential expression
$\hat D_{\alpha,\beta,\gamma,\delta,\eta}$
on proper functional spaces yield self-adjoint differential operators \cite{Ron}.
In the present article we will skip the detail, needed for
justification of the operators' domains and proper scalar products
in the corresponding linear spaces of functions.
Here we restrict our consideration only to formal manipulations
with differential expressions like (\ref{D}).

The confluent Heun's operator $\hat D_{\alpha,\beta,\gamma,\delta,\eta}$ \eqref{D} owns a remarkable property.
Its eigenfunctions $H_\lambda(z)$ for eigenvalues $\lambda\neq 0$, i.e. the solutions of the ordinary differential equation
\ben
\hat D_{\alpha,\beta,\gamma,\delta,\eta} H_\lambda(z)=\lambda H_\lambda(z)
\la{eigenfunction}
\een
are at the same time solutions of confluent Heun equation \eqref{DHeunC_} with the same parameters $\alpha, \beta, \gamma, \delta$
and a different parameter $\eta^\prime\!=\!\eta-\lambda$. For them the following confluent Heun's equation takes place
\ben
\hat D_{\alpha,\beta,\gamma,\delta,\eta-\lambda}H_\lambda(z)=0,\,\,\,\,\,\forall \lambda\in \mathbb{C}.
\la{DHeunC_lambda}
\een

Indeed, the equation \eqref{eigenfunction} obviously can be rewritten in the form \eqref{DHeunC_}
with $\mu^\prime=\mu+\lambda$ and $\nu^\prime=\nu-\lambda$. Then, using the relations
$\delta=\nu+\mu-\alpha\left({{\beta+\gamma}\over 2}-1\right)$ and
$\eta={1\over 2}(\alpha-\beta-\gamma+\alpha\beta-\beta\gamma)-\mu$
one obtains the above result.

Commuting $\hat D_{\alpha,\beta,\gamma,\delta,\eta}$ with differential expression
${\frac{d^n}{dz^n}}$ we derive the basic novel relation
\begin{align}
{\frac{d^n}{dz^n}}\hat D_{\alpha,\beta,\gamma,\delta,\eta}=
\hat D_{\alpha(n),\beta(n),\gamma(n),\delta(n),\eta(n)}{\frac{d^n}{dz^n}}+
n\,\alpha\left({\frac{\delta}{\alpha}}+{\frac{\beta+\gamma}{2}}+n\right){\frac{d^{n-1}}{dz^{n-1}}},\,\,\,n=0,1,2,\dots
\la{basic}
\end{align}
The transformation $\{\alpha, \beta, \gamma, \delta, \eta\}\to \{\alpha(n), \beta(n), \gamma(n), \delta(n), \eta(n)\}$
defined by the relations
\ben
\hskip -1.5truecm
\alpha(n)=\alpha,\, \beta(n)=\beta+n,\, \gamma(n)=\gamma+n, \,{\frac{\delta(n)}{\alpha(n)}}={\frac{\delta}{\alpha}}+n, \,
\eta(n)=\eta+{\frac n 2}(n-\alpha+\beta+\gamma)
\la{parameters_n}
\een
yields a specific augmentation of the indexes $\alpha, \beta, \gamma, \delta$ and $\eta$.

To simplify our notations we will use further on the following ones:
\begin{align}\label{notations}
\hat D_{\alpha,\beta,\gamma,\delta,\eta}\equiv\hat D_{0},\,\,\,\,\,
\hat D_{\alpha(1),\beta(1),\gamma(1),\delta(1),\eta(1)}\equiv\hat D_{1},\,\,\dots,\,\,
\hat D_{\alpha(n),\beta(n),\gamma(n),\delta(n),\eta(n)}\equiv\hat D_{n},\dots
\end{align}

Applying the equation \eqref{basic} to arbitrary solution  $H(z)$
of confluent Heun's equation \eqref{DHeunC_}
we obtain the following four-terms recurrence relation for the derivatives
${\frac{d^n}{dz^n}}H(z)=H^{(n)}(z)$:
\begin{align}
\hat D_{n}H^{(n)}(z)=-
n\,\alpha\left({\frac{\delta}{\alpha}}+{\frac{\beta+\gamma}{2}}+n\right)H^{(n-1)}(z),\,\,\,n=0,1,2,\dots
\la{dnHrecurence}
\end{align}
For $n\geq 2$ it can be considered, too, as an ordinary differential equation of third order for the functions $H^{(n-1)}(z)$.

Then applying several times the equation \eqref{dnHrecurence} we obtain the following sequence of relations
for arbitrary solution  $H(z)$ of confluent Heun's equation \eqref{DHeunC_}:
\begin{align}
\hat D_{1}\hat D_{2}\dots\hat D_{n}H^{(n)}(z)=(-\alpha)^n\,n!\left({\frac{\delta}{\alpha}}+{\frac{\beta+\gamma}{2}}+1\right)_{\!\!n}
 H(z),\,\,\,n=1,2,\dots
\la{DdnH_H}
\end{align}
Here $(x)_{n}={{\Gamma\left(x+n\right)}/
{\Gamma\left(x\right)}}=x(x+1)\dots(x+n-1)$ stands for the Pochhammer symbol \cite{BE}.

Finally, applying $\hat D_{0}$ to the both sides of equation \eqref{DdnH_H}
we end with the ordinary differential equations of order $2(n+1)$ for derivatives $H^{(n)}(z)$
of the solution  $H(z)$ of confluent Heun's equation \eqref{DHeunC_}:
\begin{align}
\hat D_{0}\hat D_{1}\dots\hat D_{n}H^{(n)}(z)=0,\,\,\,n=1,2,\dots
\la{dnHODE}
\end{align}

The relations \eqref{DdnH_H} obviously show that any solution $H(z)$ of confluent Heun's
equation \eqref{DHeunC_} is an eigenfunction of the operator
$\hat D_{1}\hat D_{2}\dots\hat D_{n}{\frac{d^n}{dz^n}}$ and
$(-\alpha)^n\,n!\left({\frac{\delta}{\alpha}}+{\frac{\beta+\gamma}{2}}+1\right)_{\!n}$ is
the corresponding eigenvalue.
The relations \eqref{dnHODE} show that the solutions $H(z)$ of the confluent Heun's equation
belong simultaneously to the null-spaces of the infinite sequence of
linear operators $\hat D_{0}\hat D_{1}\dots\hat D_{n}{\frac{d^n}{dz^n}}$: $n=0,1,2\dots$

\section{New subclass of confluent Heun's functions}

When the $\delta_N$-condition (\ref{PolynomCond:a}) is fulfilled, one obtains
${\delta}\!=\!-\alpha\left({\frac{\beta+\gamma}{2}}\!+\!N\!+\!1\right)=\delta_N$ for some fixed
nonnegative integer $N \geq 0$. Then for any function $H(z)\in \mathcal{C}^{N+3}$ the equation
(\ref{basic}) reduces to
\ben
{\frac{d^{N+1} }{dz^{N+1}}}\left(\hat D_{0} H\right)=
\hat D_{N+1}\left({\frac{d^{N+1} H}{dz^{N+1}}}\right).
\la{split}
\een

This relation shows that in the case under consideration the operator
${\frac{d^{N+1} }{dz^{N+1}}}$ defines a specific generalized Darboux transformation \cite{Takemura}
for the confluent Heun's operator $\hat D_{0}$. It is remarkable that in the right hand side of equation \eqref{split}
we have just another confluent Heun's operator -- $\hat D_{N+1}$.
As a result, if the function $H_\lambda(z)$ with $\delta=\delta_N$
is an eigenfunction of the confluent Heun operator $\hat D_{0}$ for some eigenvalue $\lambda$, then
the relation \eqref{split} shows that the function ${\frac{d^{N\!+1} }{dz^{N\!+1}}}\,H_\lambda(z)$
is an eigenfunction to the corresponding operator $\hat D_{N+1}$ for the same eigenvalue $\lambda$.

Note that the condition $\delta=\delta_N$ is invariant under arbitrary changes of the eigenvalue $\lambda$.
The operator $\hat D_{0}$ with $\delta=\delta_N$ preserves the linear envelope of
its eigenfunctions with different $\lambda$,
with the same $\delta=\delta_N$ and fixed nonnegative integer $N$.
The discussion of the equation \eqref{eigenfunction} at the beginning of the previous Section 2 shows
that studying this linear space of functions, further on we can restrict our consideration with the case $\lambda=0$.
All formulae for $\lambda\neq 0$ can be obtained from the ones for $\lambda=0$ replacing $\eta$ with $\eta-\lambda$.

Suppose the function $H(z)$ is in addition a solution of the equation
\eqref{DHeunC_}. Then the relation \eqref{split} yields the following novel confluent Heun's equation
for the derivative ${\frac{d^{N+1} H}{dz^{N+1}}}$ :
\ben
\hat D_{N+1}\left({\frac{d^{N+1} H}{dz^{N+1}}}\right)=0.
\la{HeunC_dN}
\een

Consider the two unique solutions of the Heun's differential equations (\ref{DHeunC_}) and (\ref{HeunC_dN}),
which are simultaneously regular at the point $z=0$ and equated to unity at this point.
Then these solutions are precisely the two confluent Heun's functions $\text{HeunC}(\alpha,\beta,\gamma,\delta_N,\eta,z)$
and
$\text{HeunC}\big(\alpha(N\!+1),\beta(N\!+1),\gamma(N\!+1),\delta_N(N\!+1),\eta(N\!+1), z\big)$, the last being
constructed according to equations (\ref{parameters_n}).
For them the uniqueness theorem and the uniform normalization at the point $z=0$
yield the novel relation \footnote{The explicit form the relation \eqref{HHn}  reads
\ben
\hskip -.truecm
{\frac{d^{N\!+1} }{dz^{N\!+1}}}\text{HeunC}\Bigg(\alpha,\beta,\gamma,-\alpha\left(\!{\frac{\alpha\!+\!\beta}{2}}\!+\!N\!+\!1\!\right),\eta, z\Bigg)
=(N+1)!\,v_{N\!+1}\Bigg(\alpha,\beta,\gamma,\!-\alpha\left(\!{\frac{\alpha\!+\!\beta}{2}}\!+\!N\!+\!1\!\right),\eta\Bigg)\times \nonumber \\
\hskip .truecm\times\,\text{HeunC}\Bigg(\alpha,\beta\!+\!N\!+\!1,\gamma\!+\!N\!+\!1,
-\alpha{\frac{\beta\!+\!\gamma}{2}},\eta\!+\!{\frac{N\!+1}{2}}\bigg(\!N\!+\!1\!-\!\alpha\!+\!\beta\!+\!\gamma\!\bigg),z\Bigg)
\hskip .truecm \nonumber
\een
for all $\alpha,\beta,\gamma,\eta
\in \mathbb{C}$, (when $\beta$ is not a negative integer) and for any fixed $N\in\mathbb{Z}, N\geq 0$.}
\ben
\hskip .truecm {\frac{d^{N\!+1} }{dz^{N\!+1}}}\text{HeunC}(\alpha,\beta,\gamma,\delta_N,\eta, z)=
(N+1)!\,v_{N+1}(\alpha,\beta,\gamma,\delta_N,\eta)\times \hskip 4.3truecm
\nonumber\\
\hskip 2.5truecm
\times\,\text{HeunC}\big(\alpha(N\!+1),\beta(N\!+1),\gamma(N\!+1),\delta_N(N\!+1),\eta(N\!+1), z\big).\hskip -.35truecm
\la{HHn}\een

Here we have used the $(N+1)$-th coefficient in the Taylor series expansion \eqref{HeunC}.

{\bf Definition 1:} We call {\em $\delta_N$-confluent-Heun-functions}
the functions $\text{HeunC}(\alpha,\beta,\gamma,\delta_N,\eta,z)$ which obey the $\delta_N$-condition
for some specific nonnegative integer $N\geq 0$ and denote them as
\ben
\text{HeunC}_N(\alpha,\beta,\gamma,\eta,z)=
\text{HeunC}\big(\alpha,\beta,\gamma,-\alpha\left(N+1+(\beta+\gamma)/2\right),\eta,z\big).
\la{HeunCN}
\een

{\bf Definition 2:} We call {\em associate} $\delta_N$-confluent-Heun's-functions the functions
\ben
\hskip -1.65truecm \text{HeunC}^\maltese_N(\alpha,\beta,\gamma,\eta,z)=\\
\hskip -.4truecm =\text{HeunC}\big(\alpha,\beta\!+\!N\!+\!1,\gamma\!+\!N\!+\!1,
-\alpha(\beta\!+\!\gamma)/2,\eta\!+\!(N\!+\!1)(\!N\!+\!1\!-\!\alpha\!+\!\beta\!+\!\gamma\!)/2,z\big).
\la{AHeunCN}
\een

Now the relation (\ref{HHn}) can be represented in the short form
\ben
{\frac{d^{N\!+1} }{dz^{N\!+1}}}\,\text{HeunC}_N(\alpha,\beta,\gamma,\eta,z)=
\mathcal{P}_N(\alpha,\beta,\gamma,\eta)\, \text{HeunC}^\maltese_N(\alpha,\beta,\gamma,\eta,z),
\la{HHn_maltese}
\een
where we are using the specific constant
\ben
\mathcal{P}_N(\alpha,\beta,\gamma,\eta)=(\!N\!+\!1\!)!\,v_{N\!+1}(\alpha,\beta,\gamma,-\alpha\left(N+1+(\beta+\gamma)/2\right),\eta).
\la{VN}
\een

According to the relations (\ref{parameters_n}) we obtain
${\delta_N(N+1)}/{\alpha(N+1)}+\big({\beta(N+1)+\gamma(N+1)}\big)/{2} = N+1>0$.
Thus the $\delta_N$-condition is not fulfilled for the associated
$\delta_N$-confluent-Heun's-function $\text{HeunC}^\maltese_N(\alpha,\beta,\gamma,\eta,z)$.
Hence, it does not belong to the class of the $\delta_N$-confluent-Heun-functions.

\section{A new derivation the confluent Heun's polynomials}

Now we are prepared for an alternative derivation of the confluent Heun's polynomials
without using the three-terms recurrence relation \eqref{recurrence},
i.e., directly from confluent Heun's equation.
Indeed, posing the requirement
\ben
\mathcal{P}_N(\alpha,\beta,\gamma,\eta)=0,
\la{polynomialCond_P}
\een
which is a new form of condition \eqref{PolynomCond:b},
we obtain ${\frac{d^{N+1} }{dz^{N+1}}}\text{HeunC}_N(\alpha,\beta,\gamma,\eta,z)=0$.
Thus under condition \eqref{polynomialCond_P} the
$\delta_N$-confluent-Heun's-function $\text{HeunC}_N(\alpha,\beta,\gamma,\eta,z)$
becomes a polynomial of degree $N$\footnote{Note that the condition \eqref{polynomialCond_P}
coincides with condition \eqref{PolynomCond:b} and its explicit form \eqref{Delta} up to a nonzero numerical factor.
It can be shown that the constant \eqref{VN} is simply related with Starobinsky constant
(See \cite{PF2008g} and \cite{PressTeukolsky1973}-\cite{Chandra1984}.). The equation \eqref{polynomialCond_P}
recovers the mathematical meaning of the zero-Starobinsky-constant condition.}.

The explicit form \eqref{Delta} of the polynomial condition
\eqref{polynomialCond_P} is given in the Appendix.
As seen, it presents an algebraic equation of degree $(N+1)$ for the spectral parameter $\mu$ of the confluent Heun's
equation.
Then from equation \eqref{Delta} we obtain $(N+1)$-in-number roots $\mu_{k=1,\dots,N+1}(\alpha,\beta,\gamma)$,
which yield $(N+1)$-in-number values
$\eta_{k=1,\dots,N+1}(\alpha,\beta,\gamma) =
{\frac{1} 2}(\alpha-\beta -\gamma+\alpha\beta-\beta\gamma)-\mu_{k=1,\dots,N+1}(\alpha,\beta,\gamma)$
of the parameter $\eta$.
Hence, the condition \eqref{polynomialCond_P} defines $(N+1)$-in-number polynomial solutions
\ben
\text{PHeunC}_{N,k}(\alpha,\beta,\gamma,z)=\text{HeunC}_{N}(\alpha,\beta,\gamma,\eta_k,z),\,\,\, k=1,\dots,N+1;
\la{PHeunC}
\een
to the confluent Heun's equation,
each being polynomial of degree $N$ of the variable $z$.

Further information about the mathematical properties of the confluent Heun's polynomials
$\text{PHeunC}_{N,k}(\alpha,\beta,\gamma,z)$ one can find in \cite{Ron}.
As a by-product we obtain that under conditions \eqref{PolynomCond:ab}
the confluent Heun's operator $\hat D_{\alpha,\beta,\gamma,\delta,\eta}$ \eqref{D}
becomes a quasi-solvable one in the sense of \cite{Takemura}.

The associate $\delta_N$-confluent function $\text{HeunC}^\maltese_N(\alpha,\beta,\gamma,\eta,z)$
does not become a polynomial under condition \eqref{polynomialCond_P}.
Indeed, the corresponding $\delta_N$-condition
is never fulfilled for it when $N\!\geq\! 0$ -- see the end of the previous section.
This proofs that the function $\text{HeunC}^\maltese_N(\alpha,\beta,\gamma,\eta,z)$ is not a polynomial
when the $\delta_N$-confluent function $\text{HeunC}_N(\alpha,\beta,\gamma,\eta,z)$ is.

\section{Some comments and concluding remarks}
Both for the general and for the confluent hypergeometric functions a set of
simple and universal representations of their repeated derivatives in terms of another
hypergeometric functions are well known and widely used, see for example \cite{BE1}.

At present the corresponding mathematical theory is still not developed enough
to have a complete picture for this problem in the case of different Heun's functions.
For general Heun's functions analogous relations seem to exist only for some {\em particular} cases,
see for example \cite{IsSu, Takemura, Valent}.
In the known to us literature one can not find similar relations for confluent Heun's functions.
In the present article we are filling this gap partially, taking into account the specific
properties of the confluent Heun's functions.

The general Heun's equation, written in the universal Fuchsian form:
\ben
\hskip 0.truecm
H''+\left({\frac {\gamma_{{}_G}} {z}}+{\frac{\delta_{{}_G}}{z-1}}+{\frac{\epsilon_{{}_G}}{z-1/a}}\right)H'+
{\frac{\alpha_{{}_G}\beta_{{}_G} z-q}{z(z-1)(z-1/a)}}H = 0,\,\,\,\,\,
\gamma_{{}_G}\!+\!\delta_{{}_G}\!+\!\epsilon_{{}_G}\!=\!\alpha_{{}_G}\!+\!\beta_{{}_G}\!+\!1;
\hskip .truecm
\la{HeunG}
\een
was constructed by Karl Heun in  \cite{Heun} as a generalization of the standard hypergeometric equation,
by adding one more regular singular point: $z=1/a\in (1,\infty)$.
Putting $\gamma_{{}_G}\!=\!\beta\!+\!1$, $\delta_{{}_G}\!=\!\gamma\!+\!1$, $\epsilon_{{}_G}\!=\!-\alpha/a$, $\alpha_{{}_G}\beta_{{}_G}\!=\!-(\mu\!+\!\nu)/a$, $q\!=\!-\mu/a$,
and taking the limit $a\to 0$, we obtain the confluent Heun's equation \eqref{HeunC} by coalescence of
the regular singular points $z=1/a$ and $z=\infty$ in the equation \eqref{HeunG}.

Unfortunately, such coalescence process in the very solutions and corresponding
relations is much more complicated and, as a rule, not possible.
We shall illustrate this fact using the approach of the articles
\cite{Takemura, Valent, Takemura2}.
Translating the four {\em different} regular singular points $z=0, 1, 1/a, \infty$ of the equation \eqref{HeunG}
in the compactified complex plane $\mathbb{\tilde C}_z$  to the places
of the vortexes $x=0, 1, 1+ \tau, \tau$ ($\tau$ being pure imaginary) of the fundamental rectangle of periods
in the complex plane $\mathbb{\tilde C}_x$ by the transformation
\ben
z={\frac{e_1-e_3}{\wp(x)-e_3}},\,\,\,\,\, e_{1,2,3}=\wp(\omega_{1,2,3});
\la{Weierstrass}
\een
$\wp(x)$ being Weierstrass elliptic function with half-periods
$\omega_0=0, \omega_1=1/2, \omega_2=1/2+\tau/2,\omega_3=\tau/2$,
and using in addition the substitution
\ben
\hskip .truecm
\psi(x)=z^{(l_0+1)/2}(z-1)^{(l_1+1)/2}(1-az)^{(l_2+1)/2}H(z),\,\,\,\,\,\,
l_0\!=\!\gamma_{{}_G}\!-\!3/2, l_1\!=\!\delta_{{}_G}\!-\!3/2, l_2\!=\!\epsilon_{{}_G}\!-\!3/2;
\la{H_psi}
\een
one can transform the general Heun's equation \eqref{HeunG} into elliptic form, identical to the BC1 Inozemtsev model:
\ben
-{\frac{d^2\psi(x)}{dx^2}}+\sum_{i=0}^3 l_i(l_i+1)\wp(x+\omega_i)\psi(x)=E\psi(x).
\la{Inozemtsev}
\een
Here $l_3$ and $E$ are properly chosen constants.
This equation is instrumental for deriving the results in the articles \cite{Takemura, Valent, Takemura2},
where slightly different notations are in use (See, too, the references therein.). Obviously, the equation
\eqref{Inozemtsev} has no relevant limit $a\to 0$, since $\lim_{a\to 0}\,|l_2|=\infty$.

A more fundamental {\em geometrical} obstacle to relate the results for general Heun's equation
and the results for confluent one
is that according to the formulae $e_3\!=\!-{\frac{2-a}{1+a}}e_1$, $e_2\!=\!{\frac{1-2a}{1+a}}e_1$
one obtains $\lim_{a\to 0}\,e_2\!=\!e_1$.
Hence, an elliptic representation of the confluent Heun's equation does not exists:
As a result of the coalescence process, in it we have only three different singular points.
An analytic map like \eqref{Weierstrass} of the triangle of the singular points of the confluent Heun's equation
onto a rectangle of the periods of elliptic functions is impossible.
Hence, there exist essential differences between the properties of general and confluent Heun's functions.
Note that some of the properties can be reformulated properly after the coalescence.
For example, the factor $(1-az)^{(l_2+1)/2}$ has a confluent limit $\exp(\alpha z/2)$.
As a result a correspondingly modified substitution \eqref{H_psi} exists.
It transforms the confluent Heun's equation to Schr\"odinger-like (non-elliptic) form \cite{DDMRR}-\cite{PF2009a}.

A direct consequence of the above consideration is the need to derive most of the basic properties
of confluent Heun's functions independently of the properties of general Heun's functions. Sometimes
one can use similar general methods in both cases. An example is the generalized Darboux transformation \cite{Takemura},
used in Section 3 in a specific way, and invented originally for the equation \eqref{Inozemtsev}.

One can hope that the new mathematical results for the confluent Heun's functions,
derived in the present article, will be instrumental in different analytical and numerical
applications, and especially in the relativistic theory of gravity.

\vskip .7truecm
\noindent{\bf \Large Acknowledgments}
\vskip .3truecm

The author is thankful to Denitsa~Staicova and Roumen~Borissov for numerous
discussions during the preparation of the present article.

The author is grateful to unknown referees for the useful suggestions and especially for drawing
authors attention to the reference \cite{Takemura} and to the generalized Darboux transformations,
introduced for the first time in this reference.

This article was supported by the Foundation "Theoretical and
Computational Physics and Astrophysics" and by the Bulgarian National Scientific Found
under contracts DO-1-872, DO-1-895 and DO-02-136.

\appendix

\setcounter{section}{1}

\section*{Appendix}
The confluent Heun's equation \eqref{DHeunC_} can be rewritten in the transparently self-adjoint form:

\ben
\hskip .truecm e^{-\alpha z}z^{-\beta}(z-1)^{-\gamma}{\frac{d}{dz}}
\left(e^{\alpha z}z^{1+\beta}(z-1)^{1+\gamma}{\frac{d H(z)}{dz}}\right)
+\alpha\left({\frac{\delta}{\alpha}}+{\frac{\beta+\gamma}{2}}+1\right) z H(z)=\mu H(z).
\la{Dsa}
\een

Besides, it shows that the natural spectral parameter is $\mu$.
Correspondingly, we represent the left hand side $\Delta_{N+1}(\mu)$
of condition \eqref{PolynomCond:b}
in form of the specific three-diagonal determinant:
\ben
\hskip .truecm \left|\!\!
\begin{array}{ccccccc}
\mu\!-\!q_{1}\! & 1(1\!+\!\beta) & 0 &\!\ldots\! & 0 & 0 & 0\\
N\alpha & \mu \!-\!q_{2}\!+\!1\alpha & 2(2\!+\!\beta) & \!\ldots\! & 0 & 0 & 0  \\
0 & (N\!-\!1)\alpha & \mu\!-\!q_{3}\!+\!2\alpha & \!\ldots\!& 0 & 0 & 0 \\
\vdots & \vdots & \!\vdots\! &\!\ddots\! &\vdots &\vdots &\vdots \\
0 & 0 & 0 & \!\ldots\! & \mu \!-\!q_{N\!-\!1}+(N\!-\!2)\alpha & (N\!-\!1)(N\!-\!1\!+\!\beta) & 0 \\
0 & 0 & 0 & \!\ldots\! & 2\alpha & \mu \!-\!q_{N}\!+\!(N\!-\!1)\alpha & N(N\!+\!\beta)\\
0 & 0 & 0 & \!\ldots\! & 0 & 1\alpha & \mu \!-\!q_{N\!+\!1}\!+\!N\alpha
\end{array}
\!\!\right|\!.\hskip .truecm
\la{Delta}
\een
It turns to be useful in the real calculations \cite{PF2009a}. Here $q_{n}=(n-1)(n+\beta+\gamma)$.

A similar representation of the second polynomial condition \eqref{PolynomCond:b}
in determinant form was derived in \cite{Ron}.

Note that we can develop an alternative consideration,
interchanging the places of the regular singular points $z=0$ and $z=1$.
Then because of the obvious symmetry of the equation \eqref{DHeunC} under the change
$\{\alpha,\beta,\gamma,\delta,\mu,\nu, z\}\!\to\!\{\alpha,\gamma,\beta,\delta,\nu,\mu, z-1\}$,
the spectral parameter will be $-\nu$.


\end{document}